# What Current Flows Through a Resistor?

Bob Eisenberg[1,2]

Nathan Gold[3]

Zilong Song[3]

Huaxiong Huang[1,3]

*March 22, 2019*

[1] Fields Institute for Research in Mathematical Sciences,
222 College St, Toronto, ON M5T 3J1, Canada

[2] Department of Applied Mathematics, Illinois Institute of Technology,
10 West 35th Street, Chicago, IL 60616, USA

*and*

Department of Physiology and Biophysics, Rush University,

1653 West Congress Parkway, Chicago, IL 60612, USA

[3] Department of Mathematics and Statistics, York University, Toronto, Ontario, M3J 1P3, Canada

File name: *"What current flows through a resistor; FIFTH vers March 22 2019-1.docx"*

## Abstract

Our digital technology depends on mathematics to compute current flow and design its devices. Mathematics describes current flow by an idealization, Kirchhoff's current law: all the electrons that flow into a node flow out. This idealization describes real circuits only when stray capacitances are included in the circuit design. Motivated by Maxwell's equations, we propose that current in Kirchhoff's law be defined as $\tilde{\mathbf{J}} + \varepsilon_0 \, \partial \mathbf{E}/\partial t$ where $\tilde{\mathbf{J}}$ describes **all** the movement of charge with mass, for example, the polarization of dielectrics as well as the movement of electrons. Kirchhoff's law becomes exact and universal when current is defined this way. $\tilde{\mathbf{J}} + \varepsilon_0 \, \partial \mathbf{E}/\partial t$ is the source of the magnetic field, the source of **curl B** in Maxwell's equations. Kirchhoff's laws and Maxwell's equations can use the same definition of current.

Everyone knows what flows through a resistor. Electrons flow through a resistor. Everyone knows that electrons obey Kirchhoff's current law, and all the electrons that enter a resistor, leave it. Electrons carry mass and charge as they enter or leave the resistor.

Fig. 1

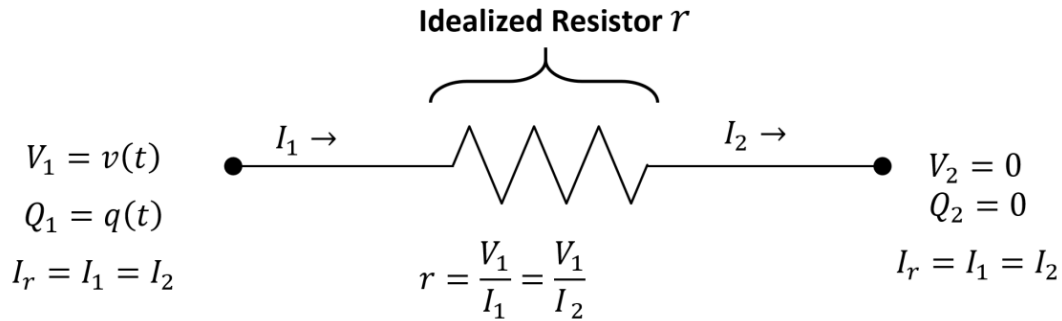

**Idealized Resistor** $r$

$V_1 = v(t)$
$Q_1 = q(t)$
$I_r = I_1 = I_2$

$r = \dfrac{V_1}{I_1} = \dfrac{V_1}{I_2}$

$V_2 = 0$
$Q_2 = 0$
$I_r = I_1 = I_2$

What everyone does not know is that this view has difficulties. All the electrons that flow in, flow out. Only electrons carry charge. So all the charge that flows in, flows out. Then, how can charge accumulate to produce the potential?

What is going on here?[1] Not much. Not much is going on if we do what engineers do, and replace the idealized resistor $r$ of Fig. 1 with a more realistic resistor $R$ that includes stray capacitance[9, 11, 16] as a separate circuit element.

Fig. 2

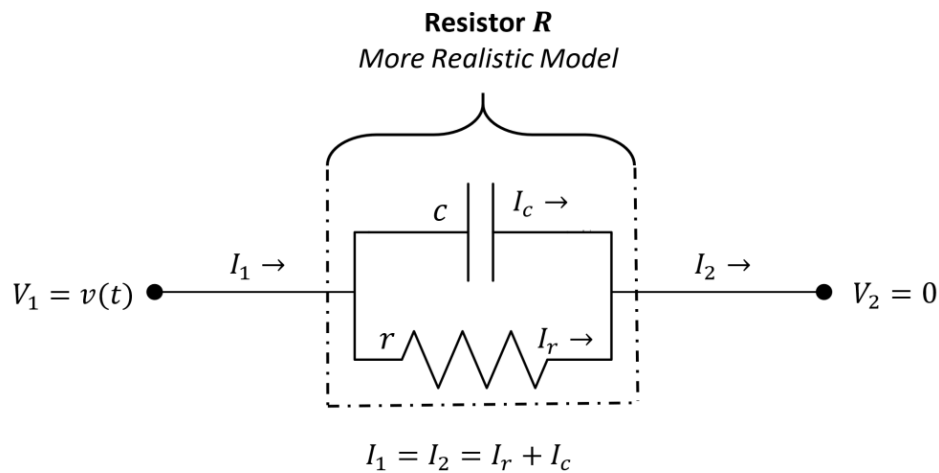

**Resistor** $R$
*More Realistic Model*

$I_1 = I_2 = I_r + I_c$

---

[1] To be precise, there are two conflicting idealizations. (1) The first idealization is *"ideal resistor $r$"*. (2) The second idealization is *"the use of Kirchhoff's current law with current equal to flux of electrons."* The conflict is traditionally resolved by changing the model of a resistor—from Fig.1 to Fig.2. An additional circuit element is added, the idealized capacitor $c$. Here we show how to resolve the conflict another way by changing the definition of current in Fig. 1, retaining the definition of the idealized resistor $r$.

The circuit in Fig. 2 is described by the circuit equations

$$I_r = \frac{v(t)}{r} \tag{1}$$

$$I_c = c\frac{\partial v(t)}{\partial t} \tag{2}$$

$$Q_1 = \int_0^T I_c dt = c\left[v(T) - v(0)\right] \tag{3}$$

$$I_1 = \frac{v(t)}{r} + c\frac{\partial v(t)}{\partial t} \tag{4}$$

We introduce an idealized parallel plate capacitor with area $A$, spacing $L$, filled with a vacuum with dielectric coefficient $\varepsilon_r = 1$ and permittivity $\varepsilon_0$ (units: farads·meter$^{-1}$).

$$c = \frac{A}{L}\varepsilon_0 \tag{5}$$

Then, we can replace the circuit equations (2)-(4) with field equations

$$I_c = \frac{A}{L}\varepsilon_0 \frac{\partial v(t)}{\partial t} \tag{6}$$

$$Q_1 = \int_0^T I_c dt = \frac{A}{L}\varepsilon_0\left[v(T) - v(0)\right] \tag{7}$$

The total current through the circuit of Fig. 2 is

$$I_1 = \frac{v(t)}{r} + \frac{A}{L}\varepsilon_0\frac{\partial v(t)}{\partial t} \tag{8}$$

In words: the more realistic resistor $\boldsymbol{R}$ of Fig. 2 has two parallel components: an idealized resistor $r$ and an idealized capacitor $c = \varepsilon_0(A/L)$. The total current through the more realistic resistor has two components. One flows through the idealized resistor $r$ defined in Fig. 1. The other component $c\,\partial v(t)/\partial t$ flows through a capacitance $c$, often called a stray or parasitic capacitance.[9, 11, 16]

In the more realistic model of a resistor (Fig. 2), there is no conflict of idealizations, because of the charge on the capacitor. All the electrons and all the charge that enters the idealized resistor $r$ leaves the resistor. But additional charge $Q_1$ flows onto the capacitor. That charge creates the electrical field.

If matter is present between the plates of the capacitor, the capacitance $c_{total}$ has two terms, one dependent on the properties of matter and the other not. If the matter is an ideal dielectric, then the capacitance of the dielectric has the vacuum component $c_0 = (A/L)\varepsilon_0$ and a material component as well.

$$c_{total} = \underbrace{\frac{A}{L}\varepsilon_0}_{c_0} + \underbrace{\frac{A}{L}(\varepsilon_r - 1)\varepsilon_0}_{Matter} \tag{9}$$

**Another description of current, motivated by Maxwell.** The charge needed to produce the potential in Fig. 2 can be described in another way if we use Maxwell's version[5, 13] of Ampere's law to define current as the source of the magnetic field.

$$\text{Maxwell's Equation:} \quad \frac{1}{\mu_o} \text{curl } \mathbf{B} = \tilde{\mathbf{J}} + \varepsilon_0 \frac{\partial \mathbf{E}(x,t)}{\partial t} \quad (10)$$

The $\varepsilon_0 \, \partial \mathbf{E}/\partial t$ term is a source of the magnetic field $\mathbf{B}$, as is $\tilde{\mathbf{J}}$. It seems natural then to define current as the entire right hand side of eq. (10).

$$\text{Definition:} \quad \textbf{Current} \triangleq \tilde{\mathbf{J}}_{\text{total}} \triangleq \tilde{\mathbf{J}} + \varepsilon_0 \frac{\partial \mathbf{E}(x,t)}{\partial t}, \quad (11)$$

Note that $\tilde{\mathbf{J}}$ includes all movements of charge, including the dielectric properties of matter. The term $\varepsilon_0 \, \partial \mathbf{E}/\partial t$ is sometimes called displacement current; it exists in a vacuum and allows light to propagate as waves in the space between stars, even though that space is a vacuum that contains (almost) zero mass[5, 13] and cannot conduct $\tilde{\mathbf{J}}$. In a vacuum like outer space, electric and magnetic fields are components of an electromagnetic wave supported by the displacement term $\varepsilon_0 \, \partial \mathbf{E}/\partial t$.

The divergence of the curl is always zero, so eq. (1) implies conservation of current:[7]

$$\text{div}\left(\tilde{\mathbf{J}} + \varepsilon_0 \frac{\partial \mathbf{E}(x,t)}{\partial t}\right) = 0 \quad (12)$$

Conservation of current in a one dimensional series circuit implies equality of current in all parts of the series circuit at all times under all conditions. The current $\tilde{\mathbf{J}}_{\text{total}} = \tilde{\mathbf{J}} + \varepsilon_0 \, \partial \mathbf{E}(x,t)/\partial t$ is equal in every component (everywhere at every time) even though the physical mechanisms producing the current differ profoundly in each component. The mechanisms can be as different as the current flow in a wire, in a semiconductor, current flow in a vacuum capacitor, and even current in an imperfect insulator. This remarkable fact is a consequence of eq. (10). Maxwell's equation (10) produces a displacement current $\varepsilon_0 \, \partial \mathbf{E}/\partial t$ that makes the total current $\tilde{\mathbf{J}} + \varepsilon_0 \, \partial \mathbf{E}/\partial t$ the same in every component in series.

We now rewrite the more realistic model of a resistor $R$ by recognizing that eq. (11) is itself the analog of eq. (4). We abandon the definition of current as the flux of electrons but we retain the definition of the ideal resistor $r$.

In equations, $\tilde{\mathbf{J}}_{\text{total}}$ is defined by eq. (11) and $r$ is defined by

$$\tilde{\mathbf{J}}_r = \frac{v(x,t)}{r} \quad (13)$$

The Maxwell-motivated definition of current eq. (11) gives the same results as the conventional circuit description Fig. 2:

$$\tilde{J}_{cap} = \varepsilon_0 \frac{\partial E(x,t)}{\partial t} = \frac{A}{L}\varepsilon_0 \frac{\partial v(x,t)}{\partial t}; \qquad c_0 = \frac{A}{L}\varepsilon_0 \tag{14}$$

**Note:** The electric field $E_{cap}(x,t)$ implied by eq. (14) is that of a parallel plate capacitor
$$E_{cap}(x,t) = \frac{A}{L}\big(v(L,t) - v(0,t)\big) \tag{15}$$

Then the total current $\tilde{J}_{total}$ through the more realistic resistor of Fig. 2 becomes

$$\tilde{J}_{total} = \tilde{J}_r + \tilde{J}_{cap} = \frac{v(x,t)}{r} + \varepsilon_0 \frac{\partial E(x,t)}{\partial t} = \frac{v(x,t)}{r} + \underbrace{\varepsilon_0 \frac{A}{L}}_{c_0} \frac{\partial v(x,t)}{\partial t} \tag{16}$$

which we recognize as eq. (4). The $\varepsilon_0\, \partial E/\partial t$ term of the Maxwell defined current $\tilde{J}_{total}$ of eq. (11) provides the current $\varepsilon_0 (A/L)(\partial v(x,t)/\partial t)$ that flows through the stray capacitance $c$ in Fig. 2.

The more realistic resistor stores charge in different ways in the circuit representation and the Maxwell formulation. In the circuit representation (of Fig. 2; eq. (1)-(8)), charge is stored in a physical element, the capacitor $c$. In the Maxwell formulation (of eq. (10)-(16)) charge is stored in the electric field itself, by the $\varepsilon_0\, \partial E/\partial t$ term. No physical capacitor is needed to store charge.[2]

**Conclusion:** the same definition of current can be used in Maxwell's equations and Kirchhoff's law.

**Our starting question** was "What current flows through a resistor?"

**Our answer.** The current that flows through an ideal resistor is $\tilde{J}_{total} = \tilde{J} + \varepsilon_0\, \partial E/\partial t$, where $\tilde{J}$ is the flux of all charge with mass.

**Our explanation.** $\tilde{J}_{total}$ is the source of the magnetic field. It is also the current that flows through the ideal resistor. The same definition of current can be used in Maxwell's equations and Kirchhoff's law.

---

[2] The additional charge storage found in capacitors filled with dielectrics can be included as it was in eq.(9). Add an additional term $(\varepsilon_r - 1)\varepsilon_0\,(A/L)$ that describes an ideal dielectric to the right hand side of eq. (16). This term is a quite poor representation of the properties of most materials and so we do not display it. We prefer to keep eq. (16) as exact as the Maxwell equations themselves.

# Additional Remarks

**Scope of paper.** This paper is meant to motivate a more general definition of current flow that removes an apparent paradox and unites Kirchhoff's and Maxwell's representations of current.

This paper is not a general analysis of real resistors. That would require a full solution of Maxwell's equations and would depend sensitively on the details of the fabrication of the resistor and how it is embedded in surroundings.[10, 11, 16]

**Redefinition of $\tilde{\mathbf{J}}$**. All movements of charge carried by or associated with mass are included in $\tilde{\mathbf{J}}$. The movements of charge include those usually described as the polarization of dielectrics, as well as more complex nonlinear polarization and other charge movements, details in [6, 7]. In many important applications, $\tilde{\mathbf{J}}$ can contain movements of charge driven by fields and forces not present in the Maxwell equations, like diffusion or convection. Nearly all of biology, and most of chemistry, occurs in electrolyte (i.e., ionic) solutions in which diffusion and convection are significant.

The other equations of Maxwell (e.g., the 'first' equation, more or less the Poisson equation) has an analogous reformulation, in which the **D** field is discarded, and the total charge (of every sort whatsoever) on the right hand side is written as explicit function(al)s of appropriate variables. Classical dielectric polarization charge would be one component; 'permanent' charge, independent of the electric field would be another; and components would appear that are driven by fields like convection, heat flow, and diffusion that are not part of classical electrodynamics at all.

The crucial point is that an explicit functional or experimental dependence needs to be specified for all the components of 'charge', just as an explicit functional or experimental dependence needs to be specified for the components of $\tilde{\mathbf{J}}$. Of course, the components of charge must spread and move in a way that produces $\tilde{\mathbf{J}}$ with its spatial spread and time dependence. $\tilde{\mathbf{J}}$ and the components of charge must be consistent, and consistent with conservation of mass and its flow, as well.

**Stray Capacitance**.[9, 11, 16] The capacitor $c$ of Fig. 2 allows a circuit model including an idealized resistor $r$ to store charge $Q_1$ as it must in any more realistic model of a resistor $R$. The capacitor is required if current is defined as the flux of electrons (or other charge carriers with mass, like ions in sea water).

The capacitor is often called parasitic or stray. It deserves to be called parasitic because the charge needed to change the potential is unavailable to flow or fan out of the (right hand terminal of) the circuit to the inputs of other devices. The capacitor is indeed a parasite because the stored charge is unavailable for other useful use. The capacitor $c$ does not deserve the name 'stray', in our opinion, because $c$ includes a component $c_0 = \varepsilon_0\ \partial\mathbf{E}/\partial t$ that cannot wander off, as strays often do.

The **un**stray component $c_0$ is the displacement current $\varepsilon_0\ \partial\mathbf{E}/\partial t$ that is the universal polarization of the vacuum also responsible for the propagation of light. The displacement current $c_0$ may be a parasite, but the parasite cannot stray.

**Special nature of the electric field**. The conservation laws of electrodynamics are exact and universal because of a special property of electrodynamics not shared by other fields, like heat flow, convection or diffusion. The vacuum polarization term $\varepsilon_0\ \partial\mathbf{E}/\partial t$ of the Maxwell equations allows the electric field to take on whatever value is necessary so the current $\tilde{\mathbf{J}}_{\text{total}}$ (of eq.(11): that includes $\varepsilon_0\ \partial\mathbf{E}/\partial t$) is exactly equal everywhere in a series circuit (see eq. (12)). No matter how mass carries charge, the displacement current $\varepsilon_0\ \partial\mathbf{E}/\partial t$ changes so the total current $\tilde{\mathbf{J}}_{total}$ —that includes the displacement current $\varepsilon_0\ \partial\mathbf{E}/\partial t$— is exactly the same everywhere in the series circuit at any time whatsoever. These issues are discussed in detail in references [7, 8].

A field like magnetism and a term like $\varepsilon_0\ \partial\mathbf{E}/\partial t$ are not found in field equations for convection, heat flow, and diffusion. The electric field can exist in a vacuum; indeed, it can 'polarize the vacuum',[3] create **curl B** through eq. (10) and thereby propagate electromagnetic waves. But convection, heat flow, and diffusion fields and flows $\tilde{\mathbf{J}}$ do not follow an equation like eq. (10) because those fields do not include an analog to displacement current. Field equations for convection, heat flow, and diffusion of the material flux $\tilde{\mathbf{J}}$ do not contain an analog to the $\varepsilon_0\ \partial\mathbf{E}/\partial t$ term in the field equations of electrodynamics that specify $\tilde{\mathbf{J}}_{total}$. The flux $\tilde{\mathbf{J}}$ describes flows of matter. Matter cannot flow or polarize in a vacuum where it does not exist. The flux $\tilde{\mathbf{J}}_{total}$ describes electrical current. Electrical current $\tilde{\mathbf{J}}_{total}$ does exist in a vacuum. The electric field polarizes a vacuum by $\varepsilon_0\ \partial\mathbf{E}/\partial t$ and thereby creates $\tilde{\mathbf{J}}_{total}$ in a vacuum devoid of matter. The electric field exists even in a vacuum without matter or its transport.

The displacement current $\varepsilon_0\ \partial\mathbf{E}/\partial t$ of the vacuum ensures that conservation of current occurs everywhere, at every time, independent of the properties of matter.[7] The flux $\tilde{\mathbf{J}}$ involves all flows of matter. The flux $\tilde{\mathbf{J}}$ can involve all the complex flows of fluid dynamics and all the flows, shock waves, and perhaps turbulence of the Navier-Stokes equations. The flux $\tilde{\mathbf{J}}$ includes the intricate movements of charge that occur within molecules and between atoms, of molecules as large and complex as proteins—with their surface layers and ionic atmospheres that move at low frequencies (~1 Hz)—and as small as organic molecules (~$10^{-17}$ sec) and even inorganic atoms. The flux occurs over time scales ranging from seconds (proteins) to say $10^{-17}$ sec of ultraviolet light (to pick one possible cut off). The movements of $\tilde{\mathbf{J}}$ are studied experimentally by various kinds of spectroscopy, with each time scale having a different name and technology: impedance,[2-4, 12] molecular,[1, 15, 17] (i.e., microwave), optical (light), ultraviolet, and electromagnetic spectroscopy (at other wavelengths) are examples.

The properties of flux of charged matter $\tilde{\mathbf{J}}$ are so diverse that general properties are hard to discern. Conservation of $\tilde{\mathbf{J}}$ — or of the corresponding flow of mass — is not apparent.

The properties of the 'current' $\tilde{\mathbf{J}}_{total}$ are much easier to grasp. Conservation is apparent,[6] as well as universal and exact in the Maxwell-Ampere equation itself, eq.(10). Conservation of electric current $\tilde{\mathbf{J}}_{total}$ depends on the mathematical identity **div curl B** $= 0$, which remains true, whatever the fluxes and properties of matter. Amazingly, this conservation law is true and useful over the entire range of behavior of the electric/magnetic field. Perhaps, that is why our digital technology—that is so fast ($10^{-9}$ sec) and yet requires near perfect reliability[14] in tiny devices made of handfuls of atoms—is built on the conservation of current in one dimensional (branched) systems. Our technology is a superstructure, even skyscraper, built on the firm foundation of Kirchhoff's law of conservation of current.

---

[3] as Maxwell called it, according to p. 228 and 416 of Darrigol[1].